\documentclass[a4paper,aps,prd,10pt,preprintnumbers,showpacs,twocolumn,superscriptaddress,nofootinbib,amsmath,amssymb]{revtex4-1}
\usepackage[dvips]{graphics}
\usepackage[utf8]{inputenc}
\usepackage[T1]{fontenc}
\usepackage{cmap}

\def\a{\widetilde{\alpha}}
\def\f{\widetilde{f}}
\def\M{{\cal M}}
\def\O{{\cal O}}

\begin{document}

\title{Simply rotating higher dimensional black holes in Einstein-Gauss-Bonnet theory}

\author{R. A. Konoplya}\email{roman.konoplya@gmail.com}
\affiliation{Research Centre for Theoretical Physics and Astrophysics, Institute of Physics, Silesian University in Opava, Bezručovo nám. 13, CZ-74601 Opava, Czech Republic}
\affiliation{Peoples Friendship University of Russia (RUDN University), 6 Miklukho-Maklaya Street, Moscow 117198, Russian Federation}

\author{A. Zhidenko}\email{olexandr.zhydenko@ufabc.edu.br}
\affiliation{Research Centre for Theoretical Physics and Astrophysics, Institute of Physics, Silesian University in Opava, Bezručovo nám. 13, CZ-74601 Opava, Czech Republic}
\affiliation{Centro de Matemática, Computação e Cognição (CMCC), Universidade Federal do ABC (UFABC), \\ Rua Abolição, CEP: 09210-180, Santo André, SP, Brazil}

\begin{abstract}
Using perturbative expansion in terms of powers of the rotation parameter $a$ we construct the axisymmetric and asymptotically flat black-hole metric in the $D$-dimensional Einstein-Gauss-Bonnet theory. In five-dimensional spacetime we find two solutions to the field equations, describing the asymptotically flat black holes, though only one of them is perturbative in mass, that is, goes over into the Minkowski spacetime when the black-hole mass goes to zero. We obtain the perturbative black-hole solution up to the order $\O(\alpha a^3)$ for any $D$, where $\alpha$ is the Gauss-Bonnet coupling, while the $D=5$ solution which is nonperturbative in mass is found in analytic form up to the order $\O(\alpha a^7)$. In order to check the convergence of the expansion in $a$ we analyze characteristics of photon orbits in this spacetime and compute frequencies of the photon orbits and radius of the photon sphere.
\end{abstract}

\pacs{04.50.-h,04.50.Kd,04.50.Gh,04.70.Bw}

\maketitle

\section{Introduction}
The problem of nonrenormalizability and existence of singularities in General relativity \cite{tHooft:1974toh,Deser:1974cz,Deser:1974xq} led to the search for viable extensions of Einstein's theory of gravity and higher-curvature corrections were taken into account \cite{Zwiebach:1985uq}. In order to avoid the Ostrogradsky instabilities \cite{Ostrogradsky:1850fid} at the classical level and ghosts in the corresponding quantum theory it is necessary to require a self-consistent model, obtained by a truncation of the higher-curvature expansion, so that the corresponding field equations still are of the second order. The most general form for such higher-curvature corrections is given by the Lovelock theory~\cite{Lovelock:1971yv,Lovelock:1972vz} which has the form of a series and the term which is quadratic in curvature is called the Gauss-Bonnet term. It does not change equations of motion in the four-dimensional spacetime, but makes nontrivial contributions in higher than four dimensions.

The first exact solution, describing spherically symmetric black holes in the higher-dimensional Einstein-Gauss-Bonnet gravity was done by D.~Boulware and S.~Deser~\cite{Boulware:1985wk}. Later their work was extended to various non-asymptotically flat and charged black holes in \cite{Wheeler,Wiltshire:1985us,Cai:2001dz}. However, all of these works were limited by nonrotating black holes and their properties.

The literature on the rotating black holes in the Einstein-Gauss-Bonnet theory, which would suggest a generalization of the higher-dimensional analogue of the Kerr metric, given by the Myers-Perry solution \cite{Myers:1986un} in the $D$-dimensional Einstein gravity, is really poor, because the corresponding field equations are complicated. Nevertheless, a few attempts to use the Kerr-Schild ansatz for finding such a generalization were made \cite{Brihaye:2008kh,Anabalon:2009kq,Ett:2011fy} and the existence of the black-hole solution in $(4+1)$-dimensional anti-de Sitter spacetime with equal angular momenta was reported, though for one specific ratio between values of the Gauss-Bonnet coupling and $\Lambda$-term \cite{Brihaye:2010wx}. The Lense-Thirring-like term describing the lowest correction of the slow rotation regime far from the $D$-dimensional Gauss-Bonnet black hole has been considered in \cite{Kim:2007iw,Adair:2020vso}.

Here we start from a rather general ansatz for the axisymmetric black hole with only one axis of rotation, which is called a \emph{simply rotating black hole}. In our opinion, this case is more interesting, because it is a more realistic model for the estimation of effects of the projection of higher-dimensional black holes onto our brane (see, for example, \cite{Kanti:2006ua,Zhidenko:2008fp} and references therein). Then, we use the perturbative procedure in the sense of an expansion in terms of the rotation parameter $a$. We obtain a black-hole solution for a generic number of spacetime dimensions $D$. However, the $D=5$ case is special: we show that within this perturbative procedure in $(4+1)$-dimensional spacetimes there are \emph{two} black-hole solutions, representing asymptotically flat black holes, which are reduced to the Myers-Perry solution \cite{Myers:1988ze} in the limit of zero Gauss-Bonnet coupling constant and to the Boulware-Deser solution \cite{Boulware:1985wk} when $a=0$. Indeed the existence of more than one black-hole solution does not contradict the uniqueness theorems, which are formulated supposing the four-dimensional gravity and absence of higher-curvature corrections.

The two obtained five-dimensional solutions are qualitatively different in a number of aspects. First of all, when the mass parameter $\M$ goes to zero, one of the solutions (which we call \emph{nonperturbative in mass}) is not reduced to the Minkowski spacetime, but diverges in this limit. This solution leads to a rather large deviation of observable quantities from their Myers-Perry limits. The other solution is perturbative in mass $\M$ and represents a rather soft deviation from the Myers-Perry geometry. The nonperturbative solution is obtained in analytic form as a series expansion up to the order $\O(\a a^7)$, while the perturbative (in mass) solution is obtained up to the order $\O(\a a^3)$ and the radial-coordinate dependence of the corresponding metric functions is calculated numerically.

In order to understand the convergence of the obtained black-hole solutions we also study some observable quantities in their background: the radius of the photon sphere and corresponding circular-orbit frequency. These are obtained both numerically and analytically as an expansion in terms of the small Gauss-Bonnet coupling constant and angular momentum.

The paper is organized as follows. In Sec.~\ref{sec:ss} we briefly summarize the properties of the spherically symmetric Boulware-Deser black hole \cite{Boulware:1985wk}. In Sec.~\ref{sec:rotating} we consider the general ansatz for the simply rotating axially symmetric black holes and the constraints we impose. There we obtain the simply rotating black-hole solution, which is perturbative in mass. The other black-hole solution, which is nonperturbative in mass, is obtained in Sec.~\ref{sec:5D}. In Sec.~\ref{sec:photon} we calculate the radius and frequency of the circular photon orbit in the background of these two black-hole metrics.
Sec.~\ref{sec:4D} is devoted to a comment on impossibility of fulfilling the $D\to4$ limit of the obtained higher dimensional formulas in order to perform the dimensional regularization in a similar fashion to \cite{Glavan:2019inb}. Finally, in Sec.~\ref{sec:discussion} we summarize the obtained results and mention some open questions.

\section{Spherically symmetric black holes in Einstein-Gauss-Bonnet theory}\label{sec:ss}
The Lagrangian density of the Einstein-Gauss-Bonnet theory has the form
\begin{eqnarray}\label{Lagrangian}
\mathcal{L} &=& R + \frac{\alpha}{2}(R_{\mu\nu\rho\sigma} R^{\mu\nu\rho\sigma} - 4 R_{\mu\nu} R^{\mu\nu} + R^2),
\end{eqnarray}
where $\alpha$ is the coupling constant, $R_{\mu\nu\rho\sigma}$ is the Riemann tensor, $R_{\mu\rho}\equiv g^{\nu\sigma}R_{\mu\nu\rho\sigma}$ is the Ricci tensor, $R\equiv g^{\mu\rho}R_{\mu\rho}$ is the Ricci scalar, and $g_{\mu\nu}$ is the $D$-dimensional metric tensor.

The Euler-Lagrange equations, corresponding to the Lagrangian density (\ref{Lagrangian}) read
\begin{eqnarray}\label{EGBequations}
R^{\mu}_{\nu}+\alpha\left(RR^{\mu}_{\nu}-2R^{\mu}_{\rho}R^{\rho}_{\nu}-2R^{\sigma}_{\rho}R^{\mu\rho}_{\phantom{\mu\rho}\nu\sigma}+R_{\nu\lambda\rho\sigma}R^{\mu\lambda\rho\sigma}\right)
&&\\\nonumber
-\frac{R}{2}\delta^{\mu}_{\nu}-\frac{\alpha}{4}\delta^{\mu}_{\nu}(R_{\rho\sigma\lambda\tau} R^{\rho\sigma\lambda\tau} - 4 R_{\rho\sigma} R^{\rho\sigma} + R^2)&=&0.
\label{Gauss-Bonnet}
\end{eqnarray}

In order to construct the metric representing a rotating black hole we start from the spherically symmetric solution and use the perturbative approach of expansion in powers of the rotation parameter $a$. Therefore, first of all, we will discuss the essentials of the spherically symmetric black-hole solution which was first obtained by D.~Boulware and S.~Deser~\cite{Boulware:1985wk}.

A spherically symmetric $D$-dimensional black hole in the Einstein-Gauss-Bonnet theory is given by the following line element
\begin{equation}\label{SSGB}
ds^2=-(1-r^2\psi(r))dt^2+\frac{dr^2}{1-r^2\psi(r)} + r^2d\Omega_{D-2}^2,
\end{equation}
where $d\Omega_{D-2}^2$ is the line element of the unit $(D-2)$-sphere.

Using the ansatz (\ref{SSGB}), the equations of the Einstein-Gauss-Bonnet theory can be reduced to the following algebraic expression for the function $\psi(r)$:
\begin{equation}\label{psi}
W[\psi(r)]\equiv\psi(r)+\a\psi^2(r)=\frac{2\M}{r^{D-1}},
\end{equation}
where $\M$ is an arbitrary constant, which defines the asymptotic mass as \cite{Myers:1988ze}
\begin{equation}\label{MADM}
M=\frac{(D-2)\pi^{D/2-3/2}}{4\Gamma(D/2-1/2)}\M.
\end{equation}

The constant $\a$ is related to the Gauss-Bonnet coupling constant $\alpha$,
\begin{equation}\label{regular}
\a=\alpha\frac{(D-3)(D-4)}{2}.
\end{equation}

When $M \to 0$, equation (\ref{psi}) has two solutions,
\begin{equation}
\psi(r)=0 \quad\mbox{and}\quad \psi(r)= - \frac{1}{\a},
\end{equation}
only the first of which represents an asymptotically flat solution. Thus, only one of the two solutions of (\ref{psi}) is perturbative in mass, that is, goes over into the Minkowski spacetime when mass goes to zero.

When $\a=0$ for the perturbative (in mass) solution of (\ref{psi}) we reproduce the Tangherlini solution \cite{Tangherlini:1963bw},
\begin{equation}\label{psi0}
\psi_0(r)=\frac{2\M}{r^{D-1}}.
\end{equation}
For $\a\neq0$, from (\ref{psi}) we can find the function $\psi(r)$, corresponding to the perturbative (in mass) solution, as a series in $\a$,
\begin{equation}\label{perturbative}
 \psi(r)=\psi_0(r)\left(1-\a\psi_0(r)+2\a^2\psi_0^2(r)-5\a^3\psi_0^3(r)+\ldots\right).
\end{equation}

\section{Rotating black hole}\label{sec:rotating}
In higher dimensions a rotating black hole may have multiple angular momenta associated with various extra dimensions. Here we are interested in the case when a single rotation occurs on our three-dimensional brane. The general form of an axisymmetric line element allows the coordinates $t$ and $\phi$ to be along the direction selected by
the two Killing vectors which are timelike and spacelike, respectively. It is convenient to choose two more spacelike coordinates, $r$ and $\theta$, to be mutually orthogonal and orthogonal to the coordinates $t$ and $\phi$, such that $r$ is the radial coordinate of the $(D-4)$-sphere, and ($r$, $\theta$, $\phi$) are spherical coordinates of the brane at spatial infinity. In this way the general form of the metric tensor for axially symmetric $D$-dimensional spacetimes with a single rotation parameter along our three-dimensional brane can be written as
\begin{eqnarray}\label{axisymmetric}
&&ds^2=-\dfrac{N^2(r,\theta)-W^2(r,\theta)\sin^2\theta}{K^2(r,\theta)}dt^2
\\\nonumber&&-2W(r,\theta)r\sin^2\theta dt \, d\phi
+\Sigma(r,\theta)\left(\dfrac{B^2(r,\theta)}{N^2(r,\theta)}dr^2 +r^2d\theta^2\right)
\\\nonumber&&+K^2(r,\theta)r^2\sin^2\theta d\phi^2+r^2\cos^2\theta d\Omega_{D-4}^2,
\end{eqnarray}
where $d\Omega_{D-4}^2$ is the line element of the unit $(D-4)$-sphere. Thus, we use a natural generalization of the four-dimensional Boyer-Lindquist coordinates. It should be noted that such a choice of coordinates completely fixes the gauge freedom for $D>4$.

When $\a=0$ we have the D-dimensional Einstein theory, and the above metric functions must have the following forms
\begin{eqnarray}\nonumber
  N^2(r,\theta)&=&1+\frac{a^2}{r^2}-r^2\psi_0(r),
\\\nonumber
  B(r,\theta)&=&1,
\\\label{MyersPerry}
  \Sigma(r,\theta)&=&1+\frac{a^2}{r^2}\cos^2\theta,
\\\nonumber
   W(r,\theta)&=&\frac{ar\psi_0(r)}{\Sigma(r,\theta)},
\\\nonumber
   K^2(r,\theta)&=&1+\frac{a^2}{r^2}+\frac{a}{r}W(r,\theta)\sin^2\theta,
\end{eqnarray}
which correspond to the particular case of the Myers-Perry black-hole solution \cite{Myers:1986un}, describing an axially symmetric D-dimensional black hole with a single rotation parameter.

It is well known that there are two kinds of black-hole instabilities:
\begin{itemize}
\item In the Einstein theory rapidly rotating black holes are unstable for large values of the rotation parameter $a$ (see \cite{Bantilan:2019bvf} and references therein) for $D\geq 6$.
\item In the Einstein-Gauss-Bonnet theory nonrotating black holes are unstable unless the coupling constant is sufficiently small \cite{Dotti:2005sq,Gleiser:2005ra,Konoplya:2008ix,Takahashi:2012np,Cuyubamba:2016cug,Konoplya:2017lhs,Konoplya:2017zwo,Konoplya:2020der}.
\end{itemize}

Therefore, it is reasonable to be limited by relatively small values of the rotation parameter $a$ and Gauss-Bonnet coupling constant $\a$, because at large $a$ and $\a$ instabilities are highly anticipated.
Thus, we will consider the perturbative solution of the rotating black hole in the Gauss-Bonnet theory in terms of two small parameters, i.~e., we study a series expansion for the metric functions with respect to $\a$ and $a^2$.

It is possible to check that, if one replaces $\psi_0(r)$ by $\psi_0(r)-\a\psi_0^2(r)$ in~(\ref{MyersPerry}),
\begin{equation}
\psi_0(r) \rightarrow \psi_0(r)-\a\psi_0^2(r)
\end{equation}
the Einstein-Gauss-Bonnet equations (\ref{EGBequations}) are satisfied as well, when neglecting the terms of order $\a^2$ and $\a a^2$. We notice that the Gauss-Bonnet corrections of the order $\a a^2$ in equations (\ref{EGBequations}) are linear combinations of the terms proportional to $\psi_0^2(r)$ and $\psi_0^2(r)\cos^2\theta$. Therefore, we choose the following ansatz for the metric functions:
\begin{eqnarray}\nonumber
  \Sigma(r,\theta)&=&1+\frac{a^2}{r^2}\cos^2\theta
\\\nonumber&&
  -\a a^2\psi_0^2(r)(f_0(r)+f_1(r)\cos^2\theta+ \O(\a,a^4)),
\\\nonumber
  N^2(r,\theta)&=&1+\frac{a^2}{r^2}-r^2\psi_0(r)+\a r^2\psi_0^2(r)
\\\nonumber&&
  +\a a^2\psi_0^2(r)(f_2(r)+f_3(r)\cos^2\theta+\O(\a,a^4)),
\\\nonumber
  B(r,\theta)&=&1-\a a^2\psi_0^2(r)(f_4(r)+f_5(r)\cos^2\theta+\O(\a,a^4)),
\\\label{GBrotation}
   W(r,\theta)&=&\frac{a}{r\Sigma(r,\theta)}(r^2\psi_0(r)-\a r^2\psi_0^2(r)
\\\nonumber&&
   -\a a^2\psi_0^2(r)(f_6(r)+f_7(r)\cos^2\theta+\O(\a,a^4))),
\\\nonumber
   K^2(r,\theta)&&=1+\frac{a^2}{r^2}+\frac{a}{r}W(r,\theta)\sin^2\theta
\\\nonumber&&
   -\a a^2\psi_0^2(r)(f_8(r)+f_9(r)\cos^2\theta+\O(\a,a^4)).
\end{eqnarray}

Substituting the ansatz (\ref{GBrotation}) into the Einstein-Gauss-Bonnet equations (\ref{EGBequations}) and considering orders up to $\O(\a a^3)$, after some calculations, we obtain equations for the ten dimensionless functions $f_i(r)$ ($i=0,1,2,3\ldots9$). Since the only dimensionless combination, which depends on $r$, is $r^2\psi_0(r)$, we define
\begin{equation}
f_i(r)=\f_i(r^2\psi_0(r))=\f_i\left(\frac{2\M}{r^{D-3}}\right),
\end{equation}
and for convenience we introduce the new radial coordinate
\begin{equation}\label{dimensionless}
x = \frac{2\M}{r^{D-3}},
\end{equation}
so that $x=0$ corresponds to spatial infinity and $x=1$ is the event horizon.

Then, the equations for the functions $\f_i(x)$ can be reduced to a system of linear differential equations with the unique solution for any $D\geq5$, such that:
\begin{itemize}
\item the corresponding metric is asymptotically flat,
\item the solution becomes the trivial one when $\M\to0$, so that the metric describes $D$-dimensional Minkowski space in this limit.
\end{itemize}
We will call the corresponding black-hole metric \emph{perturbative} in mass. In the appendix we give a detailed description of the differential equations and the numerical solution for the black-hole metric which is perturbative in mass.

The form of the above dependence of the metric on $\theta$ is justified by the fact that the ansatz (\ref{GBrotation}) is general, once we assume that the functions $\Sigma(r,\theta)$, $N^2(r,\theta)$, $B(r,\theta)$, $W(r,\theta)$, and $K^2(r,\theta)$ are analytical in $\cos\theta$, i.~e., the functions can be expanded in series of $\cos\theta$ near the equatorial plane. Indeed, if, for instance, we add terms, proportional to $\cos^4\theta$ in (\ref{GBrotation}), from the corresponding equations (\ref{EGBequations}) we find that all the coefficients in terms containing $\cos^4\theta$ vanish for the asymptotically flat metric. It is possible to check that all higher than the second powers of $\cos\theta$ in (\ref{GBrotation}) have vanishing coefficients as well, i.~e., the only nonzero coefficients are $f_i(r)$.

The event horizon $r_H(\theta)$ is given by the equation
\begin{equation}
N^2(r_H(\theta),\theta)=0,
\end{equation}
allowing us to obtain the correction of order $\O(\a a^2)$ to the shape of the horizon. Taking into account that
$$r_H(\theta)=r_0+\O(a^2)+\O(\a),$$
where $r_0$ is the black-hole radius for the Tangherlini solution, $r_0^{D-3}=2\M$, and neglecting the higher-order corrections, $\O(\a^2,\a a^4)$, we obtain the following relation,
\begin{equation}
r_H^2+a^2-\frac{2\M}{r_H^{D-5}}+\frac{4\a\M^2}{r_H^{2D-6}}+\a a^2(\f_2(1)+\f_3(1)\cos^2\theta)=0.
\end{equation}
Inspection of the differential equations for $f_{i}(x)$, which are written down explicitly in the appendix (see Eqs.~\ref{diffeqset}), shows that solutions to the equations diverge at the horizon $x=1$ unless $\f_3(1)=0$. In other words, the regularity condition at the event horizon implies that $\f_3(1)=0$ and the correction of order $\O(\a a^2)$ does not depend on $\theta$.

This way, using the general form (\ref{GBrotation}) and the numerical procedure for finding the functions $f_{i}(r)$, we have obtained the solution which describes an asymptotically flat axially symmetric black hole in $D$-dimensional spacetime and is perturbative in mass. In the next section we show that in $D=5$ spacetime there is another solution, which is nonperturbative in mass, corresponding to a different asymptotically flat black hole with single rotation. In this case the functions $f_{i}(r)$ will be found analytically.

\section{Nonuniqueness of the five-dimensional simply rotating black holes}\label{sec:5D}
In addition to the solution discussed in the previous section, which is perturbative in mass and exists for any $D\geq5$, we have found another solution for the particular case $D=5$.
This solution differs from the one discussed in the previous section in two aspects:
\begin{enumerate}
  \item It does not have the Minkowski limit when $\M \to 0$.
  \item The functions $f_{i}(x)$ can be found analytically:
\end{enumerate}
\begin{equation}\label{nonperturbativef}
\begin{array}{rcl}
\f_0(x)&=&0,
\\[8pt]
\f_1(x)&=&-\dfrac{2}{9 x^3}\left(21 x^2+35 x+60\right),
\\[8pt]
\f_2(x)&=&-\dfrac{11}{9 x^2}(x+2),
\\[8pt]
\f_3(x)&=&-\dfrac{22}{9 x^2}\left(2 x^2-x-1\right),
\\[8pt]
\f_4(x)&=&\dfrac{8}{3 x^2},
\\[8pt]
\f_5(x)&=&\dfrac{70 x+22}{9 x^2},
\\[8pt]
\f_6(x)&=&\dfrac{17 x+66}{9 x^2},
\\[8pt]
\f_7(x)&=&-\dfrac{2}{9 x^2}\left(22 x^2+37 x+99\right),
\\[8pt]
\f_8(x)&=&-\dfrac{28 x+120}{9 x^3},
\\[8pt]
\f_9(x)&=&-\dfrac{14x+14}{3 x^2}.
\end{array}
\end{equation}

The above functions $f_{i}(x)$ lead to the following metric functions, describing a family of rotating asymptotically flat five-dimensional black holes:
\begin{eqnarray}\nonumber
\Sigma(r,\theta)&=&1+\frac{a^2}{r^2}\cos^2\theta\left(1+\frac{20\a}{3\M}+\frac{70\a }{9 r^2}+\frac{28\a\M}{3 r^4}\right),
\\\nonumber
N^2(r,\theta)&=&1-\frac{2 \M}{r^2}+\frac{4\a\M^2}{r^6}+\frac{a^2}{r^2}\left(1-\frac{22\a}{9 r^2}-\frac{22\a\M}{9 r^4}\right)
\\\nonumber&&
+\frac{22\a a^2}{9r^4}\cos^2\theta\left(1+\frac{2\M}{r^2}-\frac{8\M^2}{r^4}\right),
\\\label{nonperturbative}
B(r,\theta)&=&1-\frac{8\a a^2}{3 r^4}-\frac{22\a a^2}{9r^4}\cos^2\theta\left(1+\frac{70\M}{11 r^2}\right),
\\\nonumber
W(r,\theta)&=&\frac{2\M a}{r^3\Sigma(r,\theta)}\left(1-\frac{11\a a^2}{3\M r^2}-\frac{2\a\M}{r^4}-\frac{17\a a^2}{9 r^4}\right)
\\\nonumber&&
+\frac{2\a a^3}{r^5\Sigma(r,\theta)}\cos^2\theta\left(11+\frac{74\M}{9 r^2}+\frac{88\M^2}{9 r^4}\right),
\\\nonumber
K^2(r,\theta)&=&1+\frac{a}{r}W(r,\theta)\sin^2\theta+\frac{a^2}{r^2}\left(1+\frac{20\a}{3\M}+\frac{28\a}{9 r^2}\right)
\\\nonumber&&
+\frac{14\a a^2}{3 r^4}\cos^2\theta\left(1+\frac{2\M}{r^2}\right),
\end{eqnarray}

When $a\to0$ the above expressions (\ref{nonperturbative}) approach those for the spherically symmetric Gauss-Bonnet black hole. When $\a\to0$, we obtain the simply rotating Myers-Perry black hole. However, the limit $\M\to0$ does not exist since in this case the functions $\Sigma(r,\theta)$ and $K^2(r,\theta)$ diverge.

In a similar manner we have calculated higher-order corrections in terms of the rotation parameter $a$. Higher orders in $a$ naturally lead to higher powers of $\cos\theta$. However, as the general form of the metric functions is very cumbersome, we do not write it down explicitly.\footnote{The Wolfram\textregistered{} Mathematica notebook with the expressions for the functions $\Sigma(r,\theta)$, $N^2(r,\theta)$, $B(r,\theta)$, $W(r,\theta)$, and $K^2(r,\theta)$ in their closed form up to the order $\O(\a a^7)$ is available at \url{https://arxiv.org/src/2007.10116v1/anc/GB5Dnonperturbative.nb}.} Although the solution is singular at $\M=0$, it is asymptotically flat in any order of $a$ and converges for $a^2<\M$.

For example, an expansion in orders of $1/r$ gives the following form of the metric functions:
\begin{widetext}
\begin{eqnarray}\nonumber
\Sigma(r,\theta)&=&1+\frac{a^2}{r^2}\cos^2\theta\left(1+\frac{20\a}{3\M}-\frac{29\a a^2}{3\M^2}+\frac{19\a a^4}{2\M^3}+\ldots\right)+\O\left(\frac{1}{r^4}\right),
\\\nonumber
N^2(r,\theta)&=&1-\frac{2 \M}{r^2}+\frac{a^2}{r^2}-\frac{\a a^2\sin^2\theta}{r^4}\left(\frac{22}{9}-\frac{3a^2}{4\M}-\frac{a^4}{30\M^2}+\ldots\right)+\O\left(\frac{1}{r^6}\right),
\\\label{nonperturbativea}
B(r,\theta)&=&1-\frac{\a a^2}{r^4}\left(\frac{8}{3}+\frac{22\cos^2\theta}{9}-\frac{2a^2}{3\M}-\frac{3a^2\cos^2\theta}{4\M}-\frac{a^4\cos^2\theta}{30\M^2}\ldots\right)+\O\left(\frac{1}{r^6}\right),
\\\nonumber
W(r,\theta)&=&\frac{2\M a}{r^3\Sigma(r,\theta)}\left(1-\frac{\a a^2(1-3\cos^2\theta)}{\M r^2}\left(\frac{11}{3}-\frac{59a^2}{24\M}-\frac{13a^4}{60\M^2}+\ldots\right)+\O\left(\frac{1}{r^4}\right)\right),
\\\nonumber
K^2(r,\theta)&=&1+\frac{a^2}{r^2}\left(1+\frac{20\a}{3\M}-\frac{29\a a^2}{3 r^2\M^2}+\frac{19\a a^4}{2 r^2\M^3}+\ldots\right)+\O\left(\frac{1}{r^4}\right).
\end{eqnarray}
\end{widetext}
It is evident that for $a^2<\M$ coefficients at higher-order corrections are getting smaller at each next order, indicating the convergence.

It is interesting to note that the linear (in $\a$) correction to the radius of the event horizon $r_H$ does not depend on the angular coordinate $\theta$:
\begin{eqnarray}
r_H^2&=&2\M-a^2
\\\nonumber&&-\a\left(1-\frac{5a^2}{6\M}-\frac{a^4}{12\M^2}-\frac{a^6}{24\M^3}+\ldots\right)+\O(\a^2).
\end{eqnarray}

This way, we have obtained the expansion of the metric functions up to the order $\O(\a a^3)$ for the solution which is perturbative in mass and up to the order $\O(\a a^7)$ for the solution which is nonperturbative (but still asymptotically flat for any nonzero value of the mass). Now we are in position to analyze some basic physical properties of these two solutions.

\section{Correction to the circular photon orbit}\label{sec:photon}
In order to estimate the effect due to the obtained correction we shall study the motion of a photon in the equatorial plane of the black hole (\ref{axisymmetric}) by taking $\theta=\pi/2$. Notice that, once we take $\theta=\pi/2$ in (\ref{axisymmetric}), the effective metric, in which the motion occurs, becomes $(2+1)$-dimensional and describes the equatorial plane. The equatorial plane also coincides with the one of a higher-dimensional simply rotating black-hole projected onto the $(3+1)$-dimensional brane.

The general covariant momentum of a massless particle has the form
\begin{equation}
p^\alpha \equiv \frac{dx^\alpha}{d \tau}\,,
\end{equation}
where $\tau$ is a worldline parameter. The energy $E = -p_t$ and angular momentum $L = p_\phi$ of the particle are conserved, and the null geodesic motion is described by the following ordinary differential equation for the radial coordinate:
\begin{equation}
g_{rr}\left(\frac{dr}{d \tau}\right)^2 = V_{\rm eff}(r)\,,
\end{equation}
where the effective potential is defined as \cite{Konoplya:2018arm}
\begin{eqnarray}\label{effective-potential}
&&V_{\rm eff}(r) \equiv - \left( g^{tt}E^2 - 2g^{t\phi}E L + g^{\phi\phi}L^2 \right)\Biggr|_{\theta = \pi/2}
\\\nonumber
&&=
\dfrac{K^2(r,\pi/2)}{N^2(r,\pi/2)}\left(E-\dfrac{W(r,\pi/2)}{K^2(r,\pi/2)}\dfrac{L}{r}\right)^2-\dfrac{L^2}{r^2 K^2(r,\pi/2)}\,.
\end{eqnarray}

The circular orbit corresponds to the constant value of the radial coordinate and consequently the null acceleration in the radial direction, what leads to the following conditions for the effective potential:
\begin{equation}\label{circularcond}
V_{\rm eff}(r) = 0, \qquad V_{\rm eff}'(r) = 0.
\end{equation}
\begin{widetext}
The circular-orbit frequency, which is independent of the coordinate choice, is defined as
\begin{equation}\label{frequency}
\Omega=\left|\frac{d\phi}{dt}\right|=\left|\frac{p^\phi}{p^t}\right|=\biggr|\dfrac{E~W(r,\pi/2)+L~\dfrac{N^2(r,\pi/2)-W^2(r,\pi/2)}{rK^2(r,\pi/2)}}{E~rK^2(r,\pi/2)-L~W(r,\pi/2)}\biggr|.
\end{equation}

For the solution which is perturbative in mass, by substituting (\ref{GBrotation}) into (\ref{circularcond}) we obtain the expression which allows one to find the radius of the stable photon orbit $r_{ph}$ or the photon sphere. Designating the event horizon radius of the Tangherlini black hole as $r_0^{D-3}=2\M$, we find that
\begin{eqnarray}\label{rph}
\left(\frac{r_{ph}}{r_0}\right)^{D-3}&=&\frac{1}{x_{ph}}+\frac{(D-5)a^2-2\a}{r_0^2}x_{ph}^{\frac{2}{D-3}}-\frac{\a a^2}{r_0^4}P(x_{ph})x_{ph}^{\frac{4}{D-3}}
\\\nonumber
&\pm& \frac{a\sqrt{(D-1)(D-3)}}{r_0}x_{ph}^{\frac{1}{D-3}}\Biggr(1+\frac{(D-5)(D-9)a^2+10\a}{(D-1)(D-3)r_0^2}x_{ph}^{\frac{2}{D-3}}
+\frac{\a a^2}{r_0^4}Q(x_{ph})x_{ph}^{\frac{4}{D-3}}\Biggr)+\O(\a^2,a^4),
\end{eqnarray}
where the two signs correspond to the co-rotating (for minus) and counter-rotating (for plus) orbits. Here the corrections of orders $\O(\a a^2)$ and $\O(\a a^3)$ are defined through the functions
\begin{eqnarray}
P(x)&=&\frac{2(D-3)}{(D-1)^2}\f_2'(x)+\frac{4(D-3)^2}{(D-1)^3}\f_8'(x)
+\frac{2D}{(D-1)^2}\f_2(x)+\frac{4(D-3)}{D-1}\f_8(x)+\frac{48}{(D-3)(D-1)},
\\
Q(x)&=&\frac{8 (D-3)}{(D-1)^4}\f_2''(x)+\frac{16 (D-3)^2}{(D-1)^5}\f_8''(x)
+\frac{8 (2 D-1)}{(D-1)^3}\f_2'(x)-\frac{4 (D-3)}{(D-1)^3}\f_6'(x)+\frac{8 (D-3) (4 D-5)}{(D-1)^4}\f_8'(x)
\\\nonumber&&
+\frac{2 \left(2 D^2+3 D-1\right)}{(D-3) (D-1)^2}\f_2(x)-\frac{4 D}{(D-1)^2}\f_6(x)+\frac{8 D}{(D-1)^2}\f_8(x)-\frac{7 (D-13) (D+7)}{3 (D-3)^2 (D-1)^2},
\end{eqnarray}
and the point $x_{ph}=2/(D-1)$ is related to the photon orbit in the background of the Tangherlini black hole.

\bigskip

Substituting (\ref{rph}) into (\ref{frequency}) we obtain the photon-orbit frequency:
\begin{eqnarray}
\frac{1}{\Omega}&=&r_0x_{ph}^{-\frac{1}{D-3}}\sqrt{\frac{D-1}{D-3}}\left(1+\frac{a^2(D-5)(D-1)-4\a}{2(D-3)(D-1)r_0^2}x_{ph}^{\frac{2}{D-3}}-\frac{\a a^2}{r_0^4}P_{\Omega}(x_{ph})x_{ph}^{\frac{4}{D-3}}\right)
\\\nonumber
&\pm&\frac{2a}{D-3}\left(1+\frac{2(D-5)a^2+6\a}{3(D-3)r_0^2}x_{ph}^{\frac{2}{D-3}}+\frac{\a a^2}{r_0^4}Q_{\Omega}(x_{ph})x_{ph}^{\frac{4}{D-3}}\right)+\O(\a^2,a^4),
\end{eqnarray}
where we introduced the following functions:
\begin{eqnarray}
P_{\Omega}(x)&=&\frac{2}{(D-3)(D-1)}\f_2(x)+\frac{4}{(D-1)^2}\f_8(x)+\frac{3(D+3)}{(D-3)^2(D-1)},
\\
Q_{\Omega}(x)&=&\frac{4}{(D-1)^2}\f_2'(x)+\frac{8(D-3)}{(D-1)^3}\f_8'(x)
\\\nonumber&&
+\frac{4}{D-3}\f_2(x)-\frac{2}{D-1}\f_6(x)+\frac{4(2D-3)}{(D-1)^2}\f_8(x)+\frac{32(D+1)}{3(D-3)^2(D-1)}.
\end{eqnarray}
Numerical values of $P(x_{ph})$, $Q(x_{ph})$, $P_{\Omega}(x_{ph})$, $Q_{\Omega}(x_{ph})$, and $x_{ph}$ for various $D$ are given in Table~\ref{tabl:val}.

\bigskip

In a similar manner, for the nonperturbative $5D$ solution (\ref{nonperturbativea}), we find
\begin{eqnarray}\label{rph5D}
\frac{r_{ph}^2}{2\M}&=&2-\frac{\a}{2\M}\left(1-\frac{119a^2}{12\M}+\frac{811a^4}{48\M^2}-\frac{1537a^6}{80\M^3}+\dots\right)
\\\nonumber
&\pm& a\cdot\sqrt{\frac{2}{\M}}\Biggr(1+\frac{\a}{2\M}\biggr(\frac{5}{8}-\frac{41a^2}{12\M}+\frac{311a^4}{64\M^2}
-\frac{569a^6}{120\M^3}+\dots\biggr)\Biggr)+\O(\a)^2.
\end{eqnarray}
The corresponding photon-orbit frequency is
\begin{eqnarray}\label{freq5D}
\frac{1}{\Omega\sqrt{2\M}}&=&2-\frac{\a}{8\M}\left(1-\frac{94a^2}{3\M}+\frac{905a^4}{24\M^2}-\frac{577a^6}{15\M^3}+\ldots\right)
\\\nonumber
&\pm&\frac{a}{\sqrt{2\M}}\Biggr(1+\frac{\a}{4\M}\biggr(1-\frac{31a^2}{2\M}+\frac{227a^4}{12\M^2}
-\frac{461a^6}{24\M^3}+\ldots\biggr)\Biggr)+\O(\a)^2.
\end{eqnarray}
\end{widetext}

\begin{table}
\centering
\begin{tabular}{|r|l|c|c|c|c|c|c|}
\hline
$D$&$x_{ph}$&$P(x_{ph})$&$Q(x_{ph})$&$P_{\Omega}(x_{ph})$&$Q_{\Omega}(x_{ph})$\\
\hline
 $*5$& $0.5$    & $-79.3333$ & $-27.3333$ & $-31.3333$ & $-62.0000$ \\
\hline
 $5$& $0.5$    & $8.791204$ & $9.090355$ & $1.576312$ & $4.542129$ \\
 $6$& $0.4$    & $4.227319$ & $1.964214$ & $0.630349$ & $1.905396$ \\
 $7$& $0.3333$ & $2.629768$ & $0.714790$ & $0.327762$ & $1.028113$ \\
 $8$& $0.2857$ & $1.832281$ & $0.326034$ & $0.197359$ & $0.637646$ \\
 $9$& $0.25$   & $1.366618$ & $0.169369$ & $0.130533$ & $0.432055$ \\
$10$& $0.2222$ & $1.067764$ & $0.095729$ & $0.092148$ & $0.311209$ \\
$11$& $0.2$    & $0.863101$ & $0.057327$ & $0.068238$ & $0.234422$ \\
\hline
\end{tabular}
\caption{Numerical values for the functions $P(x)$, $Q(x)$, $P_{\Omega}(x)$, $Q_{\Omega}(x)$ used for estimation of the corrections to the photon orbit for various $D$. $*$The first line corresponds to the nonperturbative five-dimensional solution for comparison.}\label{tabl:val}
\end{table}

\begin{figure*}
\centering
\resizebox{\linewidth}{!}{\includegraphics*{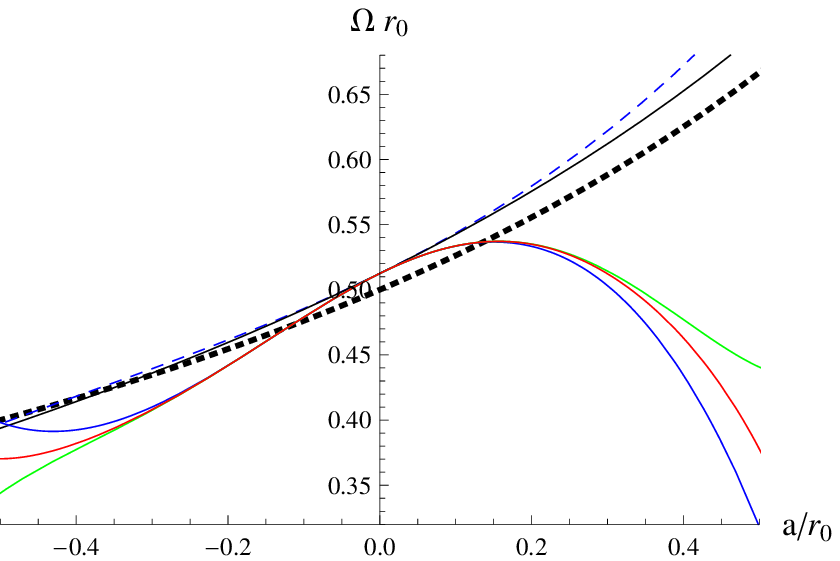}\includegraphics*{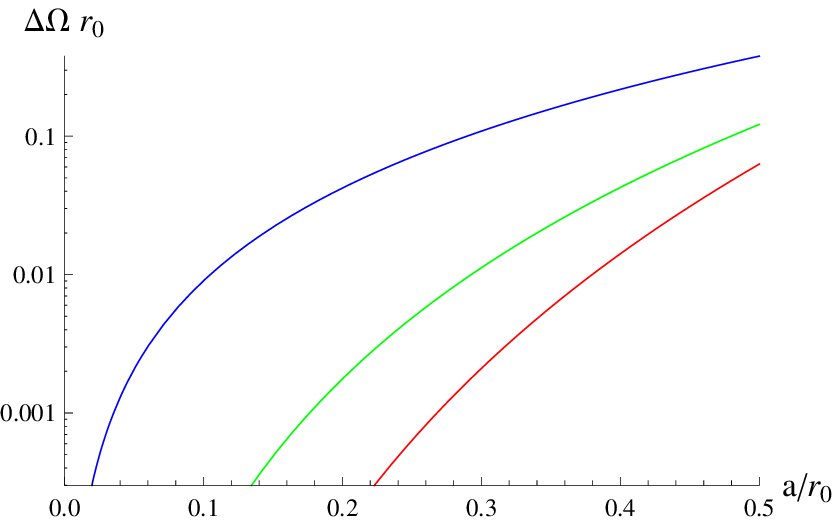}}
\caption{Left panel: circular-orbit frequency for the five-dimensional rotating Gauss-Bonnet black hole (\ref{nonperturbativea}) as a function of the rotation parameter $a$ for $\a=\alpha=0.4\M=0.2r_0$.
The black (top) line is the first-order approximation ($\O(a)$), $\Omega_1$, the blue (bottom) line is the third-order approximation ($\O(a)^3$), $\Omega_3$, the green (second form the top) line is the fifth-order approximation ($\O(a)^5$), $\Omega_5$, and the red (second from the bottom) line is the seventh-order approximation ($\O(a)^7$), $\Omega_7$. The dashed blue line corresponds to the first-order approximation for the perturbative (in mass) solution for $D=5$. The bold dotted line denotes circular-orbit frequencies of the Myers-Perry solution for comparison. A negative sign of $a$ corresponds to the counter-rotating orbits. Right panel: error estimations, $\Delta\Omega_3=|\Omega_3-\Omega_1|$ (blue, top), $\Delta\Omega_5=|\Omega_5-\Omega_3|$ (green), and $\Delta\Omega_7=|\Omega_7-\Omega_5|$ (red, bottom). Convergence becomes slower as $a$ grows. For $a^2\gtrsim\M$ the series expansion diverges.}\label{fig:freq5D}
\end{figure*}

An essential question arising when representing black-hole solutions in the form of series expansions is: what is the accuracy of the solution truncated at some order of the expansion?
The expansion has sufficient accuracy when it keeps the error being much less than the effect. For example, for given values of the Gauss-Bonnet coupling constant $\a$ and rotation parameter $a$, the difference between an observable for the Gauss-Bonnet corrected black hole and the Myers-Perry solution with the same value of the rotation parameter could be considered as the ``effect''. The order of the expected error can be estimated by the difference between observable quantities calculated for metrics truncated at various orders of the expansion. Thus, for example, from Fig.~\ref{fig:freq5D} we see that for $\a=0.4\M=0.2r_0^2$ the expansion of the seventh order in terms of the rotation parameter provides the error for the nonperturbative solution, which is one order less than the effect in the range of the rotation parameter $a\lessapprox 0.4r_0$ for both co-rotating and counter-rotating orbits. For the perturbative solutions, the co-rotating orbits are calculated with sufficient accuracy up to $a \approx 0.25r_0$, while for the counter-rotating orbits the effect is quite small as $a$ increases. The difference between the approximations of third and first order in $a$ is much smaller than the ``effect'' up to $a \approx 0.2r_0$.

We can also see that, unlike for the perturbative (in mass) solution, the photon's rotation frequency for the nonperturbative case is not a monotonous function of $a$: it first increases for the co-rotating orbit as rotation grows, reaches its maximal value $\Omega_{max}\approx0.537r_0^{-1}$ at $a\approx0.157r_0$, and then decreases. The counter-rotating orbit frequency decreases as $a$ grows within the entire parametric range, where our approximation is valid. It is interesting to notice that the corrections to the photon circular orbit have opposite signs for the perturbative and nonperturbative solutions if compared to the first-order approximation in $a$. The deviation is one order larger for the nonperturbative case (cf.~first two lines of Table~\ref{tabl:val}). Thus, the deviation of the circular-orbit frequency from their Myers-Perry value is much larger for the nonperturbative solution, while the perturbative solution stays relatively close to the Myers-Perry geometry at least when one is limited to moderate rotation (see Fig.~\ref{fig:freq5D}).

\section{On the limit $D \to 4$}\label{sec:4D}
Recently an interesting observation was made on how to construct a Gauss-Bonnet corrected solutions in the four-dimensional theory with the help of the dimensional regularization \cite{Glavan:2019inb}. Although the regularization does not form the full four-dimensional theory, in a number of cases it works as an effective tool to construction of the solutions which are also valid in the well-defined theory \cite{Aoki:2020lig}.

The system of differential equations for $f_i(r)$, which are written explicitly in the Appendix (see Eqs.~\ref{diffeqset}), as well as the corresponding solutions of these equations $f_i(r)$ do not have the limit $D\to4$. The reason is that we have used the equations corresponding to the indices on the $(D-4)$-sphere, that do not exist in the limit $D\to4$. Indeed, when considering equations of the $D$-dimensional Einstein-Gauss-Bonnet theory, after choosing a finite value of the coupling $\a$ as in (\ref{regular}), one can see that the only equations that diverge in the limit $D\to4$ are the equations with the indices of the coordinates of the $(D-4)$-sphere.

Although other equations are finite in the limit $D\to4$, we found an inconsistency in the differential equations when using the ansatz (\ref{GBrotation}). We assume that this problem may be solved in the consistent four-dimensional theory \cite{Aoki:2020lig}.

One should notice that the ansatz (\ref{GBrotation}) fixes the radial coordinate to be the radius of the $(D-4)$-sphere, which is not defined for $D=4$. Thus, there is a freedom of transformation of the coordinates, $r$ and $\theta$, allowing us to fix the function $\Sigma(r,\theta)$. In particular, the coordinate choice, such that
$$\Sigma(r,\theta)=1+\frac{a^2}{r^2}\cos^2\theta,$$
is consistent with the general ansatz proposed in~\cite{Konoplya:2016jvv}.

\section{Discussion}\label{sec:discussion}
In the present work we have obtained the following results.
\begin{itemize}
  \item We have found corrections of orders $\O(\a a^2)$ and $\O(\a a^3)$ for the metric of a simply rotating Gauss-Bonnet black hole for $D\geq5$. The corrections include a number of functions of the radial coordinate which were obtained numerically.
  \item It has been shown that for $D=5$ there exists another asymptotically flat rotating black-hole solution, which differs at $\O(\a a^2)$. This solution can be expressed analytically as a series with respect to the rotation parameter.
  \item In five-dimensional spacetime as $a\to0$ both solutions approach the same spherically symmetric Gauss-Bonnet black hole found by D.~Boulware and S.~Deser~\cite{Boulware:1985wk}. When $\a\to0$ both solutions go over into the Myers-Perry metric.
  \item As $\M\to0$ one solution approaches a flat spacetime metric, while the other one diverges.
  \item We calculated the radius and frequency of the photon's orbit for both obtained metrics and showed that the nonperturbative (in mass) solution leads to much larger deviations of observable quantities than the perturbative one.
\end{itemize}

Our work could be extended in a number of ways. First of all, it would be interesting to find a numerical solution for the simply rotating Einstein-Gauss-Bonnet black holes in order to compare the numerical solution with those obtained here by expansion in the rotation parameter. This would also allow to test the regime of fast rotation and larger coupling constants. Our approach could potentially be extended to the case of higher corrections in curvature, so the field equations would be far more complicated.

It would also be interesting to understand whether the $D=5$ case is a special one, that is, whether the nonperturbative solution exists only for $D=5$. Expansions in terms of a small rotation parameter and coupling constant indicate that for asymptotically flat black holes only the five-dimensional case allows for the nonperturbative solution. However, since we have not performed the complete analysis of the singular point at spatial infinity, we cannot rule out the possibility of the existence of the asymptotically flat nonperturbative solutions for $D>5$ or other nonperturbative solutions for $D=5$. The numerical treatment of the problem, which is not limited by small values of the parameters, could potentially answer this question.

\begin{acknowledgments}
The authors acknowledge the support of the grant 19-03950S of the Czech Science Foundation (GAČR). This publication has been prepared with partial support of the ``RUDN University Program 5-100'' (R. K.).
\end{acknowledgments}

\appendix

\section{Perturbative solution}

Substituting the ansatz into the Einstein-Gauss-Bonnet equations and considering orders $\O(\a a^2)$ and $\O(\a a^3)$, we obtain equations for the ten dimensionless functions $\f_i(x)$ ($i=0,1,2,3\ldots9$), where the dimensionless variable x is defined as (\ref{dimensionless})
$$x=\frac{2\M}{r^{D-3}},$$
so that $x=0$ corresponds to spatial infinity and $x=1$ corresponds to the Tangherlini horizon.

The functions $\f_i(x)$ satisfy the following system of linear equations:
\onecolumngrid
\begin{subequations}\label{diffeqset}
\begin{eqnarray}
\f_0(x)&=&0,
\\
\f_1''(x)&=&-\frac{5 D x-4 D-7 x+4}{(D-3) (x-1) x}\f_1'(x)
-\frac{2 \left(2 D^2 x-D^2-4 D x-D+2 x+6\right)}{(D-3)^2 (x-1) x^2}\f_1(x)
-\frac{4}{(D-3)^2 (x-1) x^2}\f_5(x)
\qquad\qquad
\\\nonumber&&
-\frac{4}{(D-3)^2 (x-1) x^2}\f_9(x)
-\frac{2 (D-1) (D+1) (D+2)}{(D-4) (D-3)^3 (x-1) x^2},
\end{eqnarray}
\begin{eqnarray}
\f_2'(x)&=&-\frac{(D x-x-2) \left(D^2 x-D^2-6 D x+4 D+6 x\right)}{2 (D-2) (2 D x-D-3 x)}\f_1'(x)
\\\nonumber&&
\!\!\!\!\!\!\!\!\!\!\!\!\!\!\!\!\!\!\!\!\!\!
-\frac{2 D^4 x^2-2 D^4 x-17 D^3 x^2+6 D^3 x+6 D^3+44 D^2 x^2+38 D^2 x-38 D^2-47 D x^2-134 D x+52 D+18 x^2+108 x}{2 (D-3) (D-2) x (2 D x-D-3 x)}\f_1(x)
\\\nonumber&&
-\frac{D+1}{(D-3) x}\f_2(x)
+\frac{D^2 x^2-3 D^2 x+2 D^2-3 D x^2+15 D x-8 D+2 x^2-14 x}{(D-3) (D-2) (x-1) x (2 D x-D-3 x)}\f_3(x)
+\frac{2}{x}\f_4(x)
\\\nonumber&&
+\frac{D x-x-2}{2 D x-D-3 x}\f_5(x)
-\frac{(D x-x-2) \left(D^2 x-3 D x-2 D+2 x\right)}{2 (D-3) (D-2) x (2 D x-D-3 x)}\f_9(x)
-\frac{(D-1) D (D+1) (D x-x-2)}{(D-3)^2 (D-2) x (2 D x-D-3 x)},
\end{eqnarray}
\begin{eqnarray}\nonumber
\f_3'(x)&=&\frac{(D-3) (D-2) (x-1) x}{2 (2 D x-D-3 x)}\f_1'(x)
+\frac{2 D^2 x-2 D^2-7 D x+8 D+5 x-8}{2 (2 D x-D-3 x)}\f_1(x)
-\frac{2 D^2 x-D^2-3 D-4 x}{(D-3) x (2 D x-D-3 x)}\f_3(x)
\\
&&
-\frac{D x^2-4 D x+2 D-x^2+4 x}{x (2 D x-D-3 x)}\f_5(x)
+\frac{D^2 x-2 D x-4 D+x+8}{2 (D-3) (2 D x-D-3 x)}\f_9(x)
-\frac{(D-1) D (D+1) (x-1)}{(D-3)^2 x (2 D x-D-3 x)},
\end{eqnarray}
\begin{eqnarray}
\f_4'(x)&=&-\frac{(D x-x-2) \left(D^2 x-D^2-6 D x+4 D+6 x\right)}{2 (D-2) (x-1) (2 D x-D-3 x)}\f_1'(x)
-\frac{(D-1) D (D+1) (D x-x-2)}{(D-3)^2 (D-2) (x-1) x (2 D x-D-3 x)}
\\\nonumber&&
\!\!\!\!\!\!\!\!\!\!\!\!\!\!\!\!\!\!\!\!\!\!
-\frac{2 D^4 x^2-2 D^4 x-17 D^3 x^2+10 D^3 x+4 D^3+44 D^2 x^2+4 D^2 x-24 D^2-47 D x^2-52 D x+32 D+18 x^2+48 x}{2 (D-3) (D-2)(x-1) x (2 D x-D-3 x)}\f_1(x)
\\\nonumber&&
+\frac{D^2 x^2-3 D^2 x+2 D^2-3 D x^2+15 D x-8 D+2 x^2-14 x}{(D-3) (D-2) (x-1)^2 x (2 D x-D-3 x)}\f_3(x)
-\frac{2 (D-1)}{(D-3) x}\f_4(x)
\\\nonumber&&
+\frac{D^2 x^2-4 D x^2+2 D x-2 D+3 x^2}{(D-3) (x-1) x (2 D x-D-3 x)}\f_5(x)
-\frac{(D x-x-2) \left(D^2 x-3 D x-2 D+2 x\right)}{2 (D-3) (D-2) (x-1) x (2 D x-D-3 x)}\f_9(x),
\end{eqnarray}
\begin{eqnarray}\nonumber
\f_5'(x)&=&\frac{(D-3) (D-2) x}{2 (2 D x-D-3 x)}\f_1'(x)
+\frac{2 D^2 x-2 D^2-7 D x+8 D+5 x-8}{2 (x-1) (2 D x-D-3 x)}\f_1(x)
-\frac{D x-2 D-x}{(D-3) (x-1) x (2 D x-D-3 x)}\f_3(x)
\\\nonumber&&
-\frac{5 D^2 x^2-6 D^2 x+2 D^2-14 D x^2+14 D x-4 D+9 x^2-6 x}{(D-3) (x-1) x (2 D x-D-3 x)}\f_5(x)
+\frac{D^2 x-2 D x-4 D+x+8}{2 (D-3) (x-1) (2 D x-D-3 x)}\f_9(x)
\\
&&
+\frac{(D-1) (D+1) \left(D^3 x-D^2 x-4 D^2+6 D x+4 D-12 x\right)}{(D-4) (D-3)^2 (D-2) (x-1) x (2 D x-D-3 x)},
\end{eqnarray}
\begin{eqnarray}
\f_6''(x)&=&\frac{(D x-x-2) \left(D^2 x-D^2-6 D x+4 D+6 x\right)}{2 (D-2) (x-1) (2 D x-D-3 x)}\f_1'(x)
-\frac{2 (2 D-1)}{(D-3) x}\f_6'(x)
\\\nonumber&&
\!\!\!\!\!\!\!\!\!\!\!\!\!\!\!\!\!\!\!\!\!\!
+\frac{2 D^4 x^2-2 D^4 x-17 D^3 x^2+10 D^3 x+4 D^3+44 D^2 x^2+4 D^2 x-24 D^2-47 D x^2-52 D x+32 D+18 x^2+48 x}{2 (D-3) (D-2)
   (x-1) x (2 D x-D-3 x)}\f_1(x)
\\\nonumber&&
-\frac{D^2 x^2+5 D^2 x-2 D^2-3 D x^2-13 D x+2 x^2+10 x}{(D-3) (D-2) (x-1)^2 x (2 D x-D-3 x)}\f_3(x)
-\frac{D^2 x^2-4 D x^2-6 D x+2 D+3 x^2+12 x}{(D-3) (x-1) x (2 D x-D-3 x)}\f_5(x)
\\\nonumber&&
-\frac{2 D (D+1)}{(D-3)^2 x^2}\f_6(x)
+\frac{2}{(D-3) (x-1) x^2}\f_7(x)
+\frac{(D x-x-2) \left(D^2 x-3 D x-2 D+2 x\right)}{2 (D-3) (D-2) (x-1) x (2 D x-D-3 x)}\f_9(x)
\\\nonumber&&
+\frac{(D-1) D (D+1) (D x-x-2)}{(D-3)^2 (D-2) (x-1) x (2 D x-D-3 x)},
\end{eqnarray}
\begin{eqnarray}
\f_7''(x)&=&-\frac{D^3 x-8 D^2 x+13 D x+4 D-6 x}{2 (D-3) (2 D x-D-3 x)}\f_1'(x)
-\frac{2 (2 D-1)}{(D-3) x}\f_7'(x)
+\frac{2 (D-1) D (D+1)^2}{(D-3)^3 (x-1) x^2 (2 D x-D-3 x)}
\\\nonumber&&
-\frac{2 D^4 x^2-2 D^4 x-19 D^3 x^2+20 D^3 x+49 D^2 x^2-50 D^2 x-8 D^2-53 D x^2+56 D x+16 D+21 x^2-24 x}{2 (D-3)^2 (x-1) x (2 D x-D-3 x)}\f_1(x)
\\\nonumber&&
+\frac{D^2 x^2+5 D^2 x-2 D^2-4 D x^2-10 D x-2 D+3 x^2+9 x}{(D-3)^2 (x-1)^2 x (2 D x-D-3 x)}\f_3(x)
\\\nonumber&&
+\frac{D^3 x^2-7 D^2 x^2-6 D^2 x+2 D^2+15 D x^2+14 D x+2 D-9 x^2-12 x}{(D-3)^2 (x-1) x (2 D x-D-3 x)}\f_5(x)
-\frac{2 (D+1) (D x-D+1)}{(D-3)^2 (x-1) x^2}\f_7(x)
\\\nonumber&&
-\frac{D^3 x^2-5 D^2 x^2-4 D^2 x+7 D x^2+4 D x+8 D-3 x^2}{2 (D-3)^2 (x-1) x (2 D x-D-3 x)}\f_9(x)
\\\nonumber&&
-\frac{(D-1)^2 (D+1) \left(5 D^3 x-6 D^3-25 D^2 x+26 D^2+22 D x+12 x-48\right)}{(D-4) (D-3)^3 (D-2) (x-1) x (2 D x-D-3 x)}
\end{eqnarray}
\begin{eqnarray}
\f_8'(x)&=&-\frac{D^2 x-D^2-6 D x+4 D+6 x}{2 D x-D-3 x}\f_1'(x)
-\frac{2 D^3 x-2 D^3-15 D^2 x+12 D^2+29 D x-14 D-18 x}{(D-3) x (2 D x-D-3 x)}\f_1(x)
\qquad\qquad
\\\nonumber&&
+\frac{2 (D x+D-2 x)}{(D-3) (x-1) x (2 D x-D-3 x)}\f_3(x)
+\frac{2 (D-2)}{2 D x-D-3 x}\f_5(x)
-\frac{2 (D-1)}{(D-3) x}\f_8(x)
\\\nonumber&&
-\frac{D^2 x-3 D x-2 D+2 x}{(D-3) x (2 D x-D-3 x)}\f_9(x)
-\frac{2 (D-1) D (D+1)}{(D-3)^2 x (2 D x-D-3 x)},
\end{eqnarray}
\begin{eqnarray}\nonumber
\f_9'(x)&=&\frac{(D-3) (D x-D-x)}{2 D x-D-3 x}\f_1'(x)
+\frac{2 D^2 x-2 D^2-5 D x+4 D+4 x}{x (2 D x-D-3 x)}\f_1(x)
-\frac{2 (D x+D-2 x)}{(D-3) (x-1) x (2 D x-D-3 x)}\f_3(x)
\\
&&
-\frac{2 (D-2)}{2 D x-D-3 x}\f_5(x)
-\frac{3 D^2 x-2 D^2-7 D x+4 D+4 x}{(D-3) x (2 D x-D-3 x)}\f_9(x)
+\frac{2 (D-1) D (D+1)}{(D-3)^2 x (2 D x-D-3 x)}.
\end{eqnarray}
\end{subequations}

Notice that, in order to have a consistent system of equations for $D>5$, we must take $\f_0(x)=0$. For the special case, $D=5$, the function satisfies
$\f_0'(x)=-4\f_0(x)/x$, and the only asymptotically flat solution is $\f_0(x)=0$.

The system of differential equations (\ref{diffeqset}) has a regular singular point at $x=0$. In order to obtain an asymptotically flat spacetime for $D>5$ one should assume that the functions $\f_i(x)$ are regular at $x=0$ ($r=\infty$). The corresponding solution is unique and perturbative in $x$ (or, equivalently, in $\M$). It can be written as the series expansion,
\begin{eqnarray}\nonumber
f_1(r)&=&\frac{D^4-3 D^3-4}{(D-4)(D-3)(D-2)^2 D}
+\frac{2\left(3D^4-3D^3+D^2+3D-4\right)}{3(D-3)(D-2)^2D(3D-7)}\left(\frac{2\M}{r^{D-3}}\right)+\O\left(\frac{2\M}{r^{D-3}}\right)^2,
\\\nonumber
f_2(r)&=&\frac{D^4-7D^3+20D^2-20D+12}{(D-4)(D-3)(D-2)^2D}
+\frac{2\left(3D^5-17D^4+29D^3+10D^2-20D+19\right)}{3(D-4)(D-3)(D-2)^2D(3D-7)}\left(\frac{2\M}{r^{D-3}}\right)+\O\left(\frac{2\M}{r^{D-3}}\right)^2,
\\\nonumber
f_3(r)&=&-\frac{D^4-3D^3-4}{(D-4)(D-3)(D-2)D}
-\frac{4 D^4-11 D^3-3 D^2+2 D-10}{3(D-4)(D-3)(D-2)^2D}\left(\frac{2\M}{r^{D-3}}\right)+\O\left(\frac{2\M}{r^{D-3}}\right)^2,
\\\nonumber
f_4(r)&=&\frac{D^2-1}{(D-4)(D-3)(D-2)D}+\frac{4 \left(D^2-1\right)^2}{3(D-4)(D-3)(D-2)^2D}\left(\frac{2\M}{r^{D-3}}\right)+\O\left(\frac{2\M}{r^{D-3}}\right)^2,
\\\label{seriesp}
f_5(r)&=&-\frac{3 D^3+D^2+2}{(D-4)(D-3)(D-2)D}
-\frac{10 D^4-8 D^3-6 D^2+8 D-4}{3 (D-4)(D-3)(D-2)^2 D}\left(\frac{2\M}{r^{D-3}}\right)+\O\left(\frac{2\M}{r^{D-3}}\right)^2,
\\\nonumber
f_6(r)&=&\frac{1}{D}+\frac{2 \left(D^3-2 D^2+2 D+5\right)}{3 (D-4) (D-2)^2 D}\left(\frac{2\M}{r^{D-3}}\right)+\O\left(\frac{2\M}{r^{D-3}}\right)^2,
\\\nonumber
f_7(r)&=&-\frac{D+1}{D-3}-\frac{D^3+D^2+2 D+2}{(D-4)(D-3)(D-2)^2 D}\left(\frac{2\M}{r^{D-3}}\right)+\O\left(\frac{2\M}{r^{D-3}}\right)^2,
\\\nonumber
f_8(r)&=&-\frac{2\left(D^3-3D^2+2D+6\right)}{(D-4)(D-3)(D-2)^2D}
\\\nonumber
f_9(r)&=&\frac{D^2+3 D+2}{(D-4)(D-3)D}+\frac{2 \left(D^3+2 D^2-D-2\right)}{3 (D-4)(D-3)(D-2)D }\left(\frac{2\M}{r^{D-3}}\right)+\O\left(\frac{2\M}{r^{D-3}}\right)^2.
\end{eqnarray}

Thus, the black-hole metric is reduced to the Minkowski space when $\M\to0$.

Further we shall consider  two additional singular points, $x=\dfrac{D}{2D-3}$ and $x=1$. In order to simplify equations (\ref{diffeqset}) we introduce the auxiliary functions,
\begin{eqnarray}\nonumber
d_0(x)&=&\frac{\f_3(x)}{(1-x)},
\\\nonumber
d_1(x)&=&\dfrac{1}{D-(2D-3)x}\Biggr(x\f_1'(x)-\dfrac{2 (3 D-5)d_0(x)}{(D-3)^2 (D-2)}
\\\label{auxfunc}
&&+\dfrac{(2 D^2-9 D+8) \f_1(x)}{(D-3) (D-2)}+\dfrac{2 \f_5(x)}{(D-3)}
-\dfrac{(D^2-7 D+8) \f_9(x)}{(D-3)^2 (D-2)}-\dfrac{2 (D^2-1) (2 D-3)}{(D-3)^3 (D-2)}\Biggr),
\\\nonumber
d_2(x)&=&x\f_6(x), \qquad
\\\nonumber
d_3(x)&=&x\f_7(x).
\end{eqnarray}
Substituting (\ref{auxfunc}) into the system of equations (\ref{diffeqset}), we obtain a system of 12 linear equations of the first order with respect to the functions $d_i(x)$ and $\f_i(x)$. The point $x=\dfrac{D}{2D-3}$ is not a singular point of the resulting system, and therefore, the function $d_1(x)$ is finite at this point.

In order to study behavior of the system at the regular singular point $x=1$, it is convenient to express the above functions $d_i(x)$ and $\f_i(x)$ in terms of linear combinations of the following functions:
\begin{eqnarray}\nonumber
d_0(x)&=& g_0(x)+\frac{D-3}{2}g_5(x)-\frac{D-2}{D-1}g(x)
-\frac{(D+1) \left(D^2-2 D+4\right)}{(D-4) (D-3) (D-2)},
\\\nonumber
d_1(x)&=& g_1(x)+\frac{4 g_1(x)}{(D-3)^3}-\frac{g_5(x)}{(D-3) (D-2)}
-\frac{2 (3 D-5) g(x)}{(D-3)^3(D-2) (D-1)}
+\frac{4 (D+1)\left(D^4-7 D^3+15 D^2-10 D-2\right)}{(D-4) (D-3)^4 (D-2)^2},
\\\nonumber
d_2(x)&=& g_2(x)-\frac{4 D - 6}{(D-2)^2}g_0(x)-\frac{(D-3)^2}{D-2}g_1(x)-g_5(x),
\\
d_3(x)&=& g_3(x)+\frac{4D-4}{(D-3)^2}g_0(x)+\frac{D+1}{D-3}g_5(x),
\\\nonumber
f_4(x)&=& g_4(x)-g_5(x)+\frac{(D-3)^2}{D-2}g_1(x)
-\frac{2 (D-1)}{(D-2)^2 (D-3)}g_0(x),
\\\nonumber
f_5(x)&=& g_5(x)+\frac{D-3}{2 (D-1)}g(x)+\frac{(D+1) \left(D^2-2 D+4\right)}{(D-4) (D-3) (D-2)},
\\\nonumber
f_9(x)&=& g(x)-g_1(x),
\end{eqnarray}
where
\begin{eqnarray}
g(x)&=&(1-x) \biggr(g_6(x)-\frac{4D-8}{(D-3)^2}g_1(x)+g_3(x)
-\frac{D-1}{2} g_5(x)+\frac{2 \left(D^3-4 D^2+8 D-7\right)}{(D-3)^2 (D-1)}\f_7(x)
\qquad\qquad\qquad\qquad\qquad\\\nonumber&&
+\frac{4 (D+1) \left(D^3-6 D+8\right)}{(D-4) (D-3)^3 (D-2)}\biggr)+\f_7(x).
\end{eqnarray}

In terms of the new functions the system of equations (\ref{diffeqset}) in the vicinity of the singular point $x=1$ takes the following form:
\begin{equation}
  \begin{array}{lll}
g_0'(x)=\dfrac{g_0(x)}{1-x}+\O(1),
\qquad&
g_1'(x)=\dfrac{g_1(x)}{1-x}+\O(1),\qquad
\\
g_2'(x)=\O(1),
&
g_3'(x)=\O(1),
&
g_4'(x)=\O(1),
\\
g_5'(x)=\dfrac{D-1}{D-3}\dfrac{g_5(x)}{1-x}+\O(1),
\qquad&
g_6'(x)=\dfrac{g_6(x)}{1-x}+\O(1),\qquad
\\
f_1'(x)=\O(1),
&
f_2'(x)=\O(1),
\\
f_6'(x)=\O(1),
&
f_7'(x)=\O(1),
&
f_8'(x)=\O(1).
  \end{array}
\end{equation}
If the solution is finite at the horizon ($x=1$), then $g_5(1)$ must be finite. 
We performed numerical integration with the initial conditions at $x=x_0\ll1$ obtained using the series expansion (\ref{seriesp}), which allows us to obtain the metric functions numerically \cite{NotebokReferece} for $x_0\leq x<1$. However, numerical integration procedures for the initial value problem, such as Runge-Kutta methods, become numerically unstable as $x\to1$, because $g_5(x)$ diverges at this point for whatever small numerical error. We observe that, for $x\to1$, there is no convergence of the solution as we increase numerical precision and accuracy of integration. We conclude that, in order to calculate the near-horizon corrections to the metric functions, one should employ other algorithms, such as finite difference methods etc.

Although the direct integration fails near the event horizon, we believe that the metric functions, obtained numerically in this way, are accurate for the whole space except for the near-horizon region, where we cannot neglect the higher-order corrections due to rotation and the coupling $\a$. In particular, using this numerical solution, we can study corrections of the order $\O(\a a^2)$ and $\O(\a a^3)$ to the radiation processes and particle orbits.


\begin{thebibliography}{99}
\bibitem{tHooft:1974toh}
  G.~'t Hooft and M.~J.~G.~Veltman,
  Ann.\ Inst.\ H.\ Poincare Phys.\ Theor.\ A {\bf 20}, 69 (1974).

\bibitem{Deser:1974cz}
  S.~Deser and P.~van Nieuwenhuizen,
  Phys.\ Rev.\ D {\bf 10}, 401 (1974)
  doi:10.1103/PhysRevD.10.401.

\bibitem{Deser:1974xq}
  S.~Deser, H.~S.~Tsao and P.~van Nieuwenhuizen,
  Phys.\ Rev.\ D {\bf 10}, 3337 (1974)
  doi:10.1103/PhysRevD.10.3337.

\bibitem{Zwiebach:1985uq}
  B.~Zwiebach,
  Phys.\ Lett.\  {\bf 156B}, 315 (1985)
  doi:10.1016/0370-2693(85)91616-8.

\bibitem{Ostrogradsky:1850fid}
  M.~Ostrogradsky,
  Mem.\ Acad.\ St.\ Petersbourg {\bf 6}, no. 4, 385 (1850).

\bibitem{Lovelock:1971yv}
  D.~Lovelock,
  J.\ Math.\ Phys.\  {\bf 12}, 498 (1971)
  doi:10.1063/1.1665613.

\bibitem{Lovelock:1972vz}
  D.~Lovelock,
  J.\ Math.\ Phys.\  {\bf 13}, 874 (1972)
  doi:10.1063/1.1666069.

\bibitem{Boulware:1985wk}
  D.~G.~Boulware and S.~Deser,
  Phys.\ Rev.\ Lett.\  {\bf 55}, 2656 (1985)
  doi:10.1103/PhysRevLett.55.2656.

\bibitem{Wheeler}
  J.~T.~Wheeler,
  Nucl.\ Phys.\ B {\bf 273}, 732 (1986)
 doi:10.1016/0550-3213(86)90388-3;
  Nucl.\ Phys.\ B {\bf 268}, 737 (1986)
 doi:10.1016/0550-3213(86)90268-3.

\bibitem{Wiltshire:1985us}
  D.~L.~Wiltshire,
  Phys.\ Lett.\  {\bf 169B}, 36 (1986)
  doi:10.1016/0370-2693(86)90681-7.

 \bibitem{Cai:2001dz}
   R.~G.~Cai,
  Phys.\ Rev.\ D {\bf 65}, 084014 (2002)
  doi:10.1103/PhysRevD.65.084014
  [hep-th/0109133].

\bibitem{Myers:1986un}
  R.~C.~Myers and M.~J.~Perry,
  Annals Phys.\  {\bf 172}, 304 (1986)
  doi:10.1016/0003-4916(86)90186-7.

\bibitem{Brihaye:2008kh}
  Y.~Brihaye and E.~Radu,
  Phys.\ Lett.\ B {\bf 661}, 167 (2008)
  doi:10.1016/j.physletb.2008.02.005
  [arXiv:0801.1021 [hep-th]].

\bibitem{Anabalon:2009kq}
  A.~Anabalon, N.~Deruelle, Y.~Morisawa, J.~Oliva, M.~Sasaki, D.~Tempo and R.~Troncoso,
  Class.\ Quant.\ Grav.\  {\bf 26}, 065002 (2009)
  doi:10.1088/0264-9381/26/6/065002
  [arXiv:0812.3194 [hep-th]].

\bibitem{Ett:2011fy}
  B.~Ett and D.~Kastor,
  JHEP {\bf 1104}, 109 (2011)
  doi:10.1007/JHEP04(2011)109
  [arXiv:1103.3182 [hep-th]].

\bibitem{Brihaye:2010wx}
  Y.~Brihaye, B.~Kleihaus, J.~Kunz and E.~Radu,
  JHEP {\bf 1011}, 098 (2010)
  doi:10.1007/JHEP11(2010)098
  [arXiv:1010.0860 [hep-th]].

\bibitem{Kim:2007iw}
H.~C.~Kim and R.~G.~Cai,
Phys. Rev. D \textbf{77}, 024045 (2008)
doi:10.1103/PhysRevD.77.024045
[arXiv:0711.0885 [hep-th]].

\bibitem{Adair:2020vso}
C.~Adair, P.~Bueno, P.~A.~Cano, R.~A.~Hennigar and R.~B.~Mann,
Phys. Rev. D \textbf{102}, no.8, 084001 (2020)
doi:10.1103/PhysRevD.102.084001
[arXiv:2004.09598 [gr-qc]].

\bibitem{Kanti:2006ua}
  P.~Kanti, R.~A.~Konoplya and A.~Zhidenko,
  Phys.\ Rev.\ D {\bf 74}, 064008 (2006)
  doi:10.1103/PhysRevD.74.064008
  [gr-qc/0607048].

\bibitem{Zhidenko:2008fp}
  A.~Zhidenko,
  Phys.\ Rev.\ D {\bf 78}, 024007 (2008)
  doi:10.1103/PhysRevD.78.024007
  [arXiv:0802.2262 [gr-qc]].

\bibitem{Myers:1988ze}
  R.~C.~Myers and J.~Z.~Simon,
  Phys.\ Rev.\ D {\bf 38}, 2434 (1988)
  doi:10.1103/PhysRevD.38.2434.

\bibitem{Glavan:2019inb}
  D.~Glavan and C.~Lin,
  Phys.\ Rev.\ Lett.\  {\bf 124}, no. 8, 081301 (2020)
  doi:10.1103/PhysRevLett.124.081301
  [arXiv:1905.03601 [gr-qc]].

\bibitem{Tangherlini:1963bw}
  F.~R.~Tangherlini,
  Nuovo Cim.\  {\bf 27}, 636 (1963).

\bibitem{Bantilan:2019bvf}
  H.~Bantilan, P.~Figueras, M.~Kunesch and R.~Panosso Macedo,
  Phys.\ Rev.\ D {\bf 100}, no. 8, 086014 (2019)
  doi:10.1103/PhysRevD.100.086014
  [arXiv:1906.10696 [hep-th]].

\bibitem{Dotti:2005sq}
  G.~Dotti and R.~J.~Gleiser,
  Phys.\ Rev.\ D {\bf 72}, 044018 (2005)
  doi:10.1103/PhysRevD.72.044018
  [gr-qc/0503117].

\bibitem{Gleiser:2005ra}
  R.~J.~Gleiser and G.~Dotti,
  Phys.\ Rev.\ D {\bf 72}, 124002 (2005)
  doi:10.1103/PhysRevD.72.124002
  [gr-qc/0510069].

\bibitem{Konoplya:2008ix}
  R.~A.~Konoplya and A.~Zhidenko,
  Phys.\ Rev.\ D {\bf 77}, 104004 (2008)
  doi:10.1103/PhysRevD.77.104004
  [arXiv:0802.0267 [hep-th]].

\bibitem{Takahashi:2012np}
  T.~Takahashi,
  PTEP {\bf 2013}, 013E02 (2013)
  doi:10.1093/ptep/pts049
  [arXiv:1209.2867 [gr-qc]].

\bibitem{Cuyubamba:2016cug}
  M.~A.~Cuyubamba, R.~A.~Konoplya and A.~Zhidenko,
  Phys.\ Rev.\ D {\bf 93}, no. 10, 104053 (2016)
  doi:10.1103/PhysRevD.93.104053
  [arXiv:1604.03604 [gr-qc]].

\bibitem{Konoplya:2017lhs}
  R.~A.~Konoplya and A.~Zhidenko,
  JCAP {\bf 1705}, 050 (2017)
  doi:10.1088/1475-7516/2017/05/050
  [arXiv:1705.01656 [hep-th]].

\bibitem{Konoplya:2017zwo}
  R.~A.~Konoplya and A.~Zhidenko,
  JHEP {\bf 1709}, 139 (2017)
  doi:10.1007/JHEP09(2017)139
  [arXiv:1705.07732 [hep-th]].

\bibitem{Konoplya:2020der}
R.~A.~Konoplya and A.~Zhidenko,
Phys.\ Lett.\ B \textbf{807}, 135607 (2020)
doi:10.1016/j.physletb.2020.135607
[arXiv:2005.02225 [gr-qc]].

\bibitem{Konoplya:2018arm}
R.~Konoplya, Z.~Stuchlík and A.~Zhidenko,
Phys. Rev. D \textbf{97}, no.8, 084044 (2018)
doi:10.1103/PhysRevD.97.084044
[arXiv:1801.07195 [gr-qc]].

\bibitem{Aoki:2020lig}
K.~Aoki, M.~A.~Gorji and S.~Mukohyama,
Phys.\ Lett.\ B {\bf 810}, 135843 (2020)
doi:10.1016/j.physletb.2020.135843
[arXiv:2005.03859 [gr-qc]];
JCAP \textbf{09}, 014 (2020)
doi:10.1088/1475-7516/2020/09/014
[arXiv:2005.08428 [gr-qc]].

\bibitem{Konoplya:2016jvv}
R.~Konoplya, L.~Rezzolla and A.~Zhidenko,
Phys. Rev. D \textbf{93}, no.6, 064015 (2016)
doi:10.1103/PhysRevD.93.064015
[arXiv:1602.02378 [gr-qc]].

\bibitem{NotebokReferece}
The Wolfram Mathematica\textregistered{} code for numerical integration of the differential equations is available at \mbox{\url{https://arxiv.org/src/2007.10116v1/anc/GBnumerical.nb}.}

\end{thebibliography}
\end{document}